\documentclass{llncs}
\usepackage{makeidx}  
\usepackage{graphicx}
\usepackage[latin1]{inputenc}
\begin{document}
\frontmatter          
\pagestyle{headings}  

\mainmatter              

\title{Data Mining on Crash Simulation Data}
\titlerunning{Crash Simulation Data}

\author{Annette Kuhlmann\inst{1} \and Ralf--Michael Vetter\inst{1} \and Christoph Lübbing\inst{2} \and Clemens--August Thole\inst{1}}
\authorrunning{Kuhlmann, Vetter et al.}   

\institute{Fraunhofer Institute for Algorithms and Scientific Computing (SCAI),\\ Schloss Birlinghoven, 53754 Sankt Augustin, Germany\\
\email{\{kuhlmann, vetter, thole\}@scai.fraunhofer.de}
\and
BMW AG, EK-210, Knorrstrasse 147, 80788 München, Germany
\email{christoph.luebbing@bmw.de}}

\maketitle 

\begin{abstract}
The work presented in this paper is part of the cooperative research project AUTO--OPT 
carried out by twelve partners from the automotive industries. One major work package 
concerns the application of data mining methods in the area of automotive design. 
Suitable methods for data preparation and data analysis are developed. 
The objective of the work is the re--use of data stored in the crash--simulation 
department at BMW in order to gain deeper insight into the interrelations between 
the geometric variations of the car during its design and its performance in crash testing. 
In this paper a method for data analysis of finite element models and results from 
crash simulation is proposed and application to recent data from the industrial partner 
BMW is demonstrated. All necessary steps from data pre--processing to re--integration 
into the working environment of the engineer are covered.
\end{abstract}

\section{Introduction}

The objective of the data mining work presented in this paper
is the re--use of data stored in the crash--simulation department at BMW in order 
to gain deeper insight into the interrelations. Here the objective is to find 
hidden know\-ledge in stored data. In principle one could think of various 
possible questions for such a knowledge mining analysis:
\begin{itemize}
\item which innovations have evolved during the design process
\item were certain steps in the development unnecessary or could they be shortened
\item is it possible to extract analogies between different car projects
\item can reasons that have lead to certain design decisions be reproduced
\item can this reasoning be applied to future projects
\end{itemize}
The data mining project in AUTO--OPT aims at examining the applicability of data 
mining methods on crash simulation data \cite{autoopt}. Due to the fact that design and development knowledge 
is the major asset of engineering, an automotive company cannot be expected to share large amounts of their data
for research reasons. On the other hand, interesting results from data mining can only be achieved from interesting data.
Therefore in this work the applicability of the method is demonstrated, its value cannot be evaluated on the data basis available.
This will be aimed at in future work.

\begin{figure}[h]
\centering \includegraphics[width=12.0cm,angle=0]{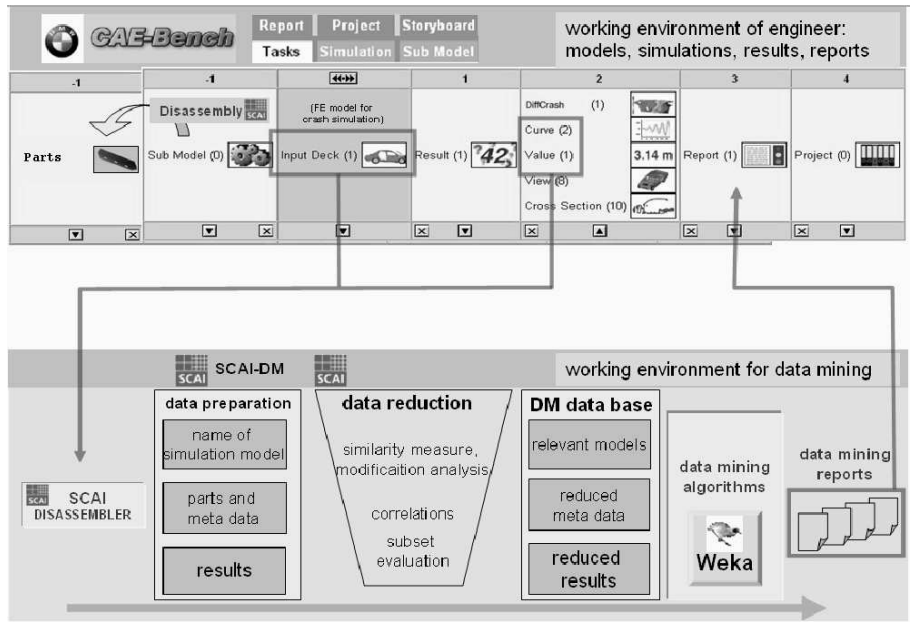}
\caption{Procedure for data mining. Models and crash results are exported 
from CAE--Bench, models are disassembled into parts, geometry based meta 
data are calculated, parts and meta data are stored within SCAI--DM, 
crash results are attached to meta data, data mining tables are assembled, 
DM analysis is performed, result files are produced and exported.}
\label{procedure_for_data_mining}
\end{figure}

The crash department at BMW stores all relevant information in the simulation data management system 
CAE--Bench \cite{caebench}. Data mining queries are to be submitted from 
this environment. Results have to be brought back into this system and assigned 
to the underlying models, i.e. stored within their audit trail. This procedure is 
schematically shown in Figure~\ref{procedure_for_data_mining}.

\begin{figure}[h]
\begin{minipage}[c]{0.5\textwidth}
\centering \includegraphics[width=1.8cm,angle=270]{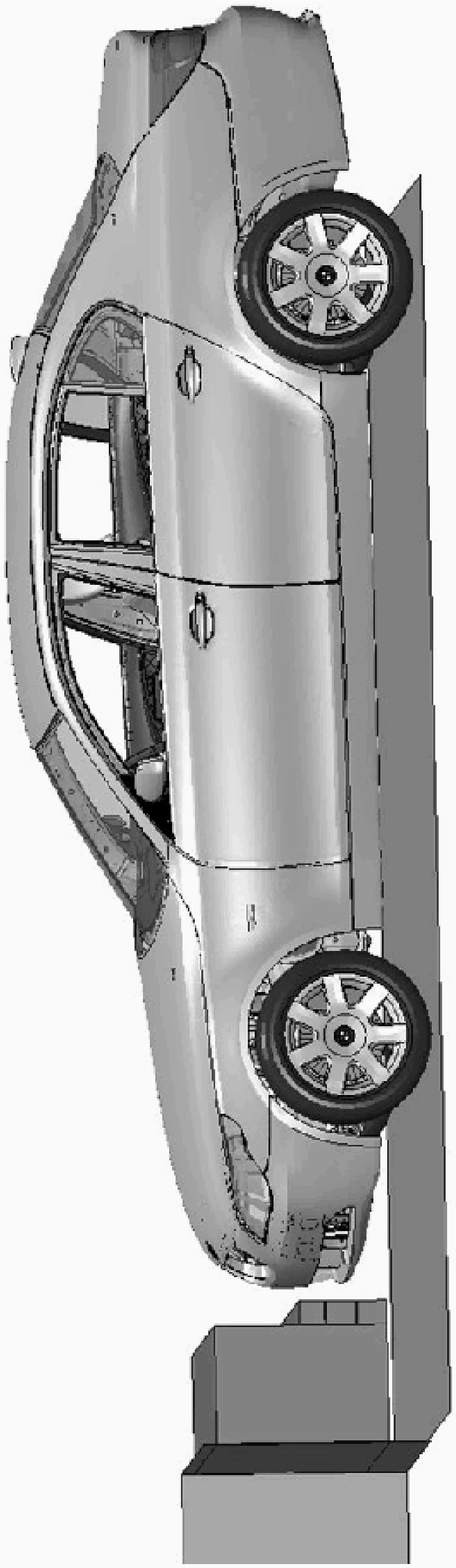}
\end{minipage}
\begin{minipage}[c]{0.5\textwidth}
\centering \includegraphics[width=3.3cm,angle=270]{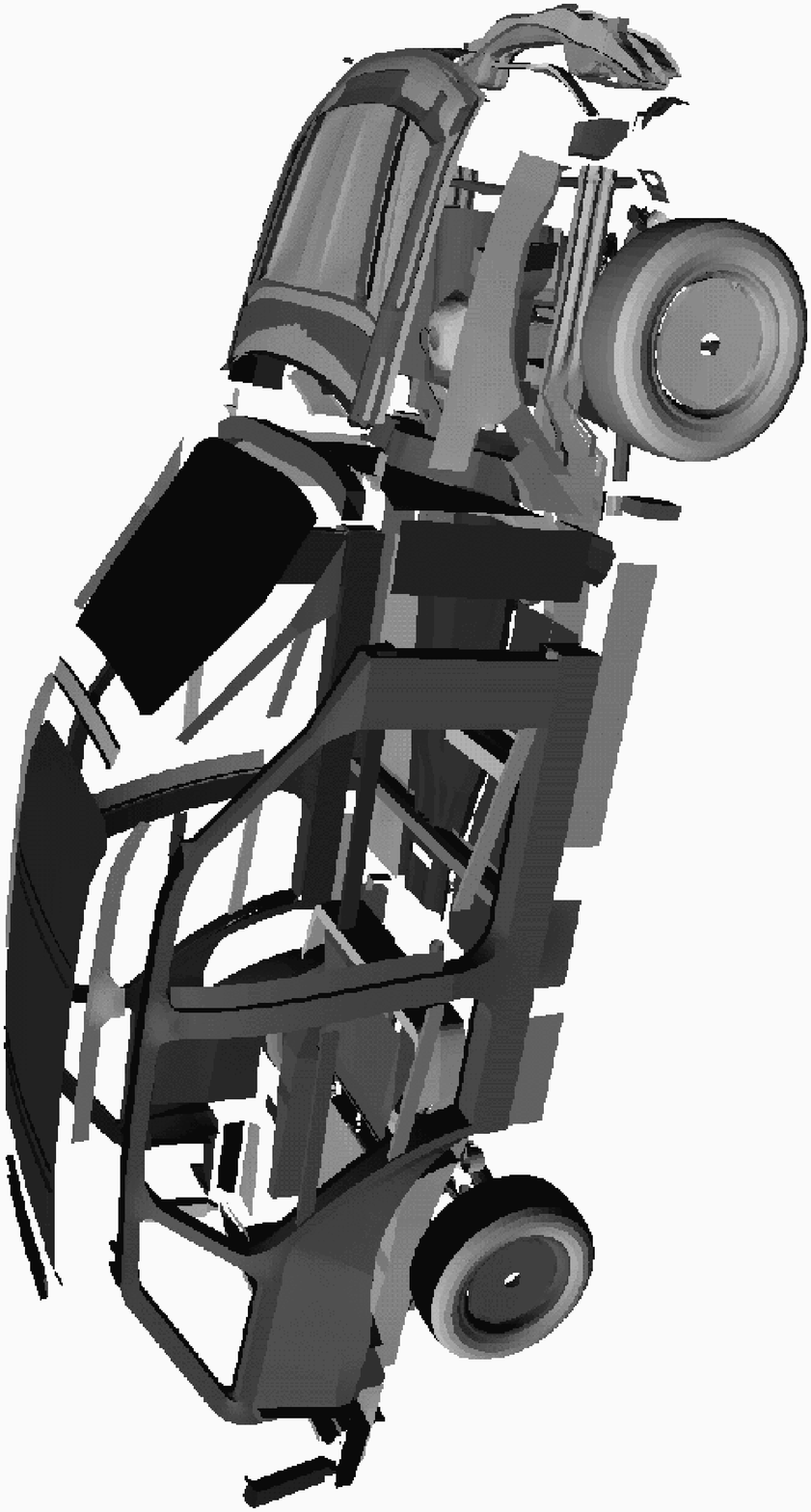}
\end{minipage}
\caption{Recent model of BMW employed for data mining (left). One FE input deck consisting of numerous parts (right).}\label{bmw_models}
\end{figure}

Fraunhofer SCAI has been provided with data from one of the most recent car 
projects at BMW. The vehicle under development is shown in Figure~\ref{bmw_models} (left). 
Each data set describes one stage of construction of this car within the 
development process via finite element models (FE models) made up by 
about 500.000 independent nodes and elements. 

Each car is composed of app. 1200 parts. CAE--Bench stores the models 
as complete vehicles, i.e. one single large FE model, called input deck. 
In order to analyse the geometry of the parts these input decks need to 
be disassembled as shown in Fig.~\ref{bmw_models} (right) for an older 
BMW vehicle, the new model cannot be shown in such detail because of a 
nondisclosure agreement.

\section{Preparation of the data for data mining}

It is generally accepted that the preparation of the data involves as much as 80-90\% of 
the effort when a data mining task is attempted, see e.g. \cite{handbook_of_data_mining}. 
The data cannot be processed by a data mining tool in their original format. To the authors
best knowledge no approach for data mining on raw finite element data exists. 
The preparation of the data thus constitutes the main challenge for the data mining approach on the FE data. 
In addition, the data has to be cleaned and checked for consistency and the appropriate 
values have to be combined. As a first but major step a process for data preparation has been developed:
\begin{itemize}
\item[{\bf a)}] Export of data from CAE--Bench
\item[{\bf b)}] Disassembling into parts and computation of meta data
\item[{\bf c)}] Data cleaning and sorting --- clustering of parts
\item[{\bf d)}] Similarity analysis and data reduction --- clustering of variants
\item[{\bf e)}] Evaluation and cleaning of crash result data
\end{itemize}
As a result of this procedure a table is generated that allows for access to the data with 
data mining algorithms. This section focuses on the preparation of the data, 
whereas the application of the data mining algorithms will be presented in section~\ref{part_ii}.

\subsection*{a) Export of data from CAE-Bench}

CAE--Bench can export selected input decks along with the result 
achieved when these models were subjected to a virtual crash test. 
An example is shown in Fig.~\ref{excerpt_of_an_input_deck}. 
Information is extracted from this export, such that the relevant crash results 
can be attached to the respective input deck data and stored in the SCAI data 
mining framework (SCAI--DM).

\subsection*{b) Disassembling into parts and computation of meta data}

\subsubsection*{Motivation.}

\begin{figure}[h]
\begin{minipage}[c]{0.35\textwidth}
\centering \includegraphics[width=0.54\textwidth,angle=0]{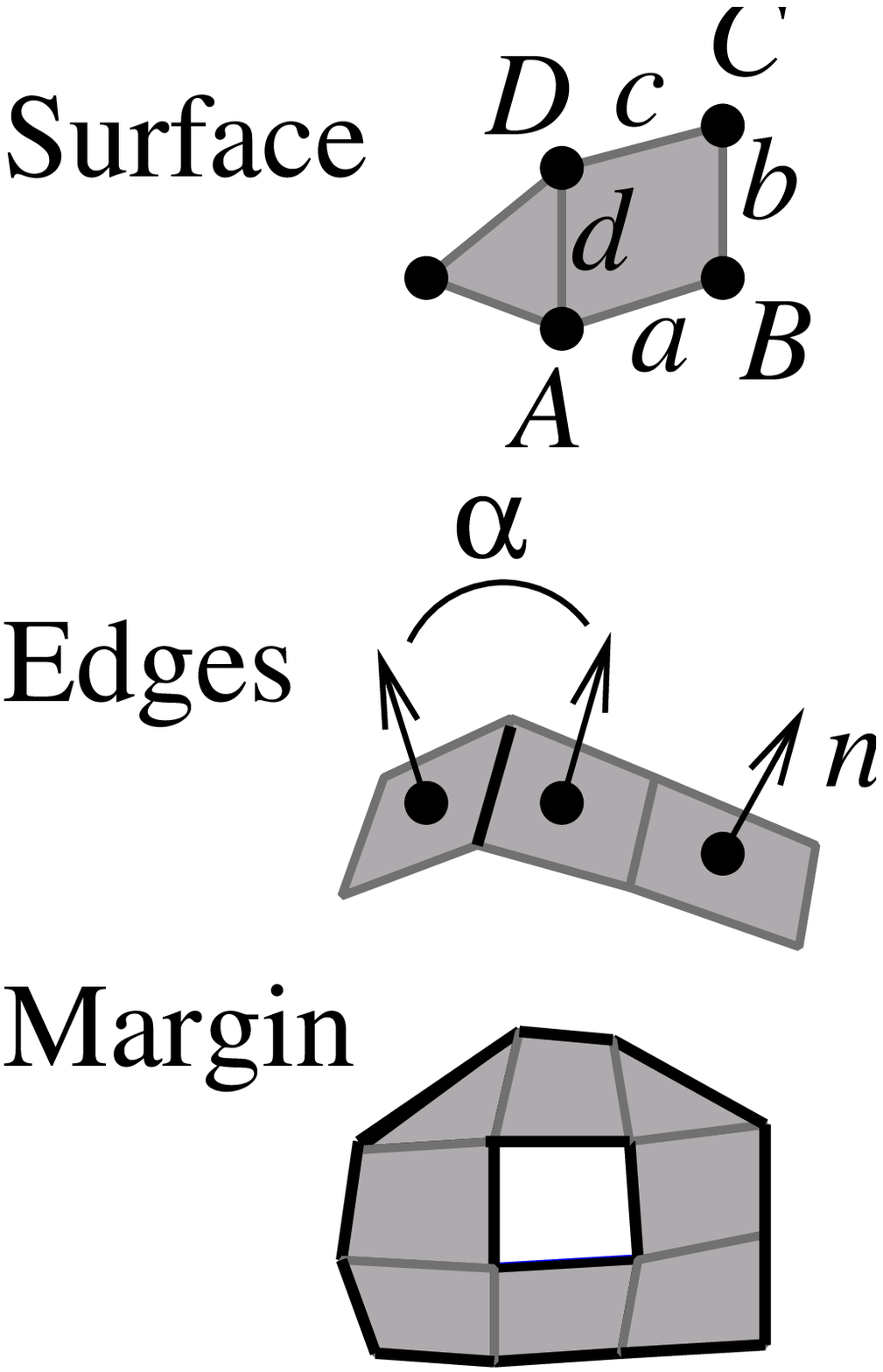}
\end{minipage}
\begin{minipage}[c]{0.68\textwidth}
\centering \includegraphics[width=0.62\textwidth,angle=270]{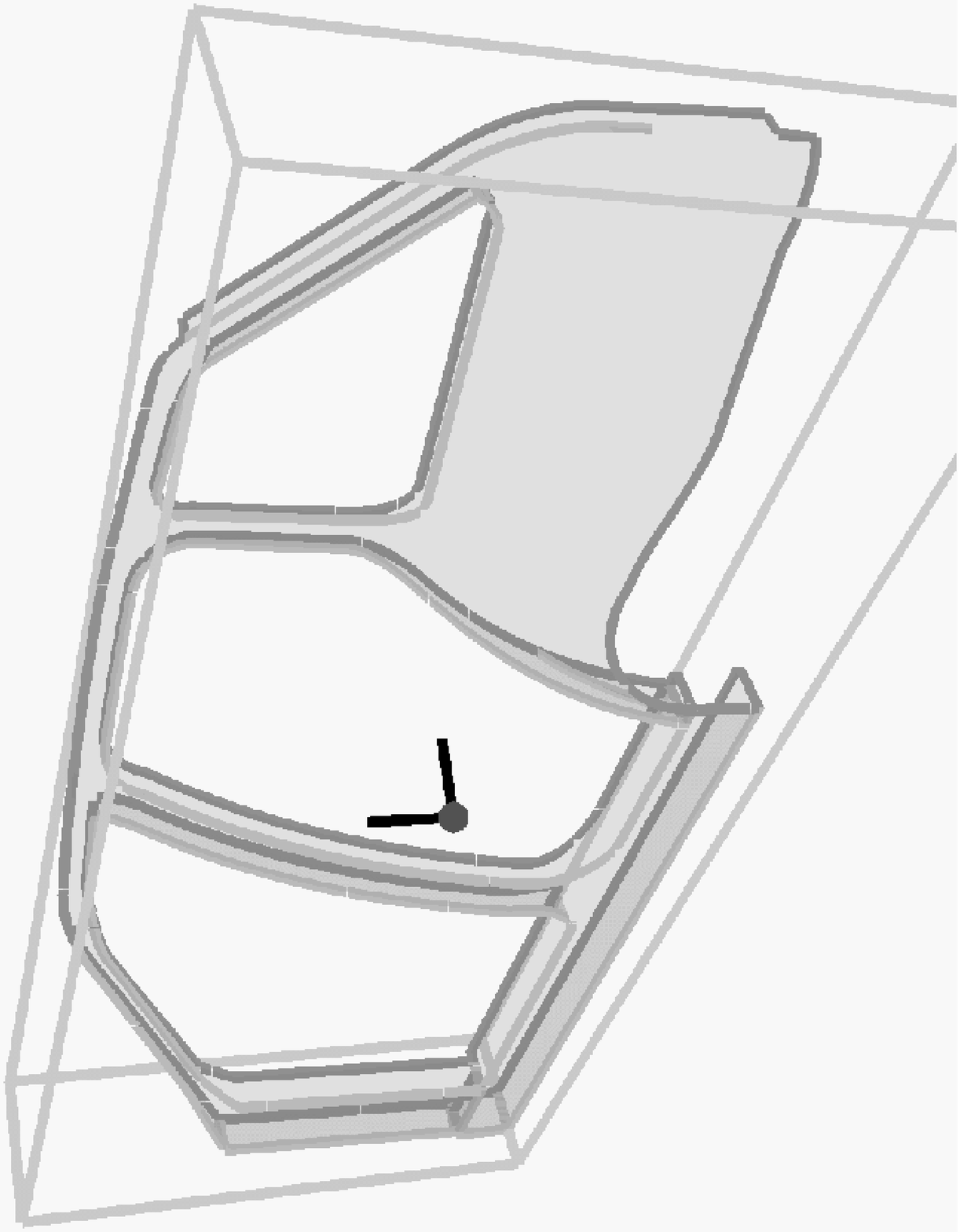}
\end{minipage}
\caption{Typical meta data and their appearance in an example part.}
\label{one_part}
\end{figure}

The data mining approach in this work concerns the shape of the parts of the car. 
The aim is to analyse how changes in shape have influenced crash behaviour. 
The FE--model itself contains all geometrical information. However, this 
information is hidden from data mining algorithms, as these cannot extract 
meaningful knowledge from node and element descriptions. Meta data has to be 
determined such that it quantifies geometry in an appropriate manner. In this 
work several values have been chosen as meta data, e.g. the centre of gravity 
of each part, the moments of inertia, the length of edges and margins, surface size, bounding box, 
length of branching lines---as shown in Figure~\ref{one_part}. All of these are 
mesh independent. They thus enable comparison between models that have been 
meshed with different algorithms or programs. The meta data reduce the 
amount of data massively, such that handling of data is facilitated considerably. 

\subsubsection*{Reading the Input Deck.}

\begin{figure}[h]
\centering \includegraphics[width=10.2cm,angle=0]{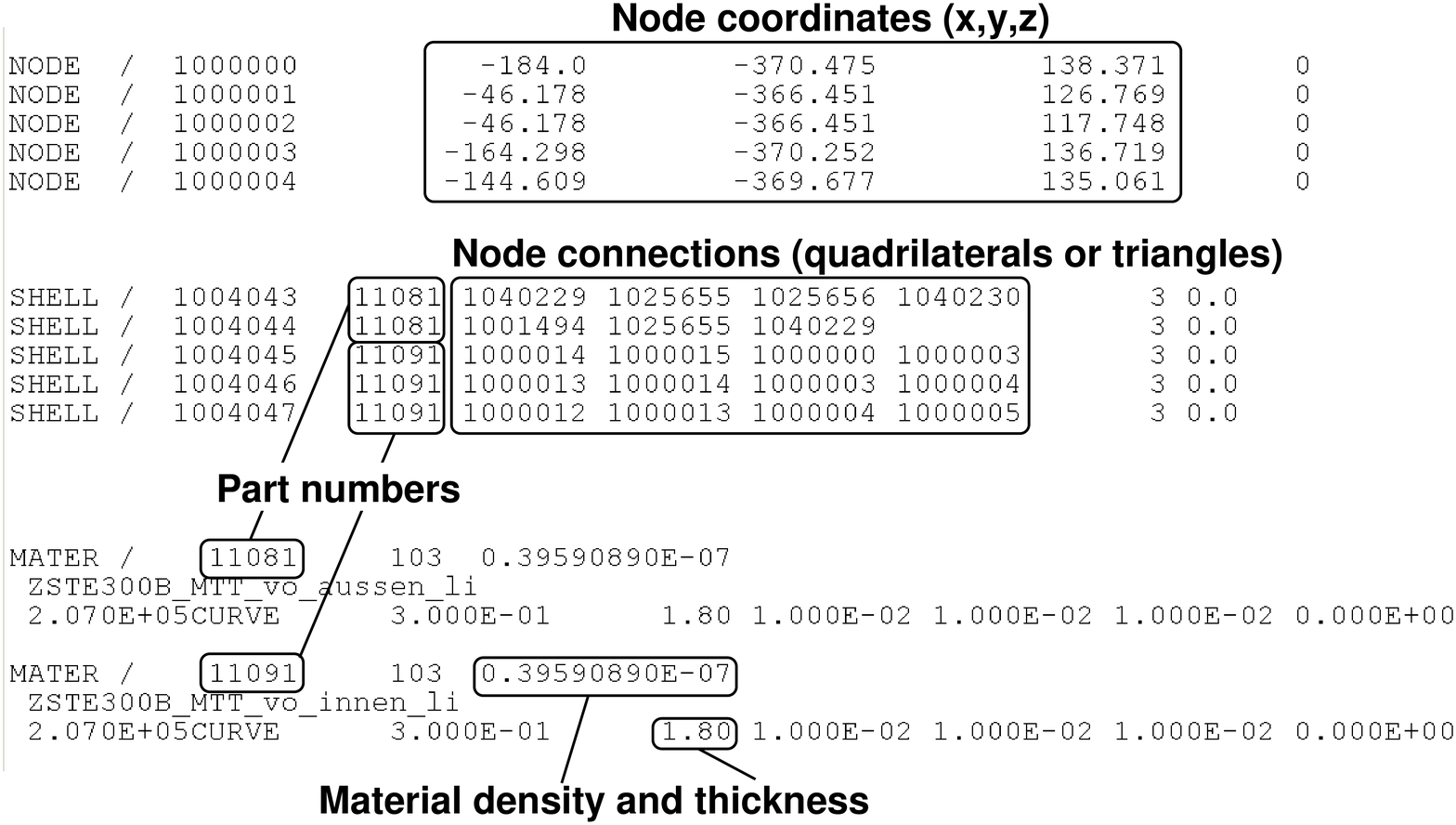}
\caption{Excerpt of an input deck: the NODE and SHELL sections describe the geometry 
of the finite element meshes; in the MATER section the thickness $d$ and density $\rho$ 
of the material can be found. Beside shell elements, which can be triangles or 
quadrilaterals, additional elements like membrane elements (with 4 nodes), 
solid (8 nodes), beams (2 nodes), or bars (3 nodes) appear in an input deck.}
\label{excerpt_of_an_input_deck}
\end{figure}

Today the body--shell of a finite element car model is described by an input deck 
of 100~MB containing approximately 1.500.000 lines. Figure~\ref{excerpt_of_an_input_deck} 
shows a small subset of such an input deck. The part 
number indicates which element of the meshes belongs to which specific part 
of the car. The material section defines a homogeneous density and thickness 
for each part. In the disassembling procedure all elements with the same part 
number and their respective nodes are extracted from the input deck and form 
one new mesh for this single part.

For each part the disassembler thus extracts a sub--mesh of the input deck. 
This sub--mesh is the basis of the calculation of the meta data. The sub--mesh files are 
also used to create previews of the parts: from three different angles or---on demand---in 
form of a three--dimensional applet visualisation \cite{javaview,javaview2,keim}. 
Since the generation of previews of parts is a time intensive process, it is initiated 
only for new parts which have not previously been stored in the database.

\subsubsection*{Computation of Meta Data.}

For meta data calculation various details on FE models have to be taken into account. 
The model surfaces here are curved. Using shell elements for the description means 
that the four corner points of a quadrilateral do not necessarily lie in one common plain, see Fig~\ref{one_part}. 
One well defined way to calculate their 
surface is $S \approx \frac{1}{2} \left[ ({\bf a}+{\bf c}) \times ({\bf b}+{\bf d}) \right]$. 
The mass of an element is given by its surface multiplied by the material thickness $d$ and 
density $\rho$. The centre of gravity, at which the mass $m$ is assumed to be located 
in a single element is positioned approximately at $\frac{1}{4} \left[ {\bf A} + {\bf B} + {\bf C} + {\bf D} \right]$, 
where ${\bf A}$ \ldots ${\bf D}$ are the corner points of the SHELL element. Then the centre of gravity 
and the moments of inertia of the complete part are given by their sum over all point masses. 
For every element a normal vector is constructed by ${\bf n}=\frac{{\bf a}+{\bf c}}{|{\bf a}+{\bf c}|} \times \frac{{\bf b}+{\bf d}}{|{\bf b}+{\bf d}|}$
The normal vectors ${\bf n_1}$ and ${\bf n_2}$ of adjacent elements are used for detecting edges. If the angle
$\alpha = 2 \arcsin\left(|{\bf n_2} - {\bf n_1}|/2\right)$ is larger than a user defined value, the connection line between the elements is called an edge of 
the mesh. If one side of an element is not connected to any further element, this line is assumed 
to be a margin of the structure. In this manner all meta data characterising the geometry of each part is computed.

\begin{figure}[h]
\centering \includegraphics[width=7.5cm,angle=270]{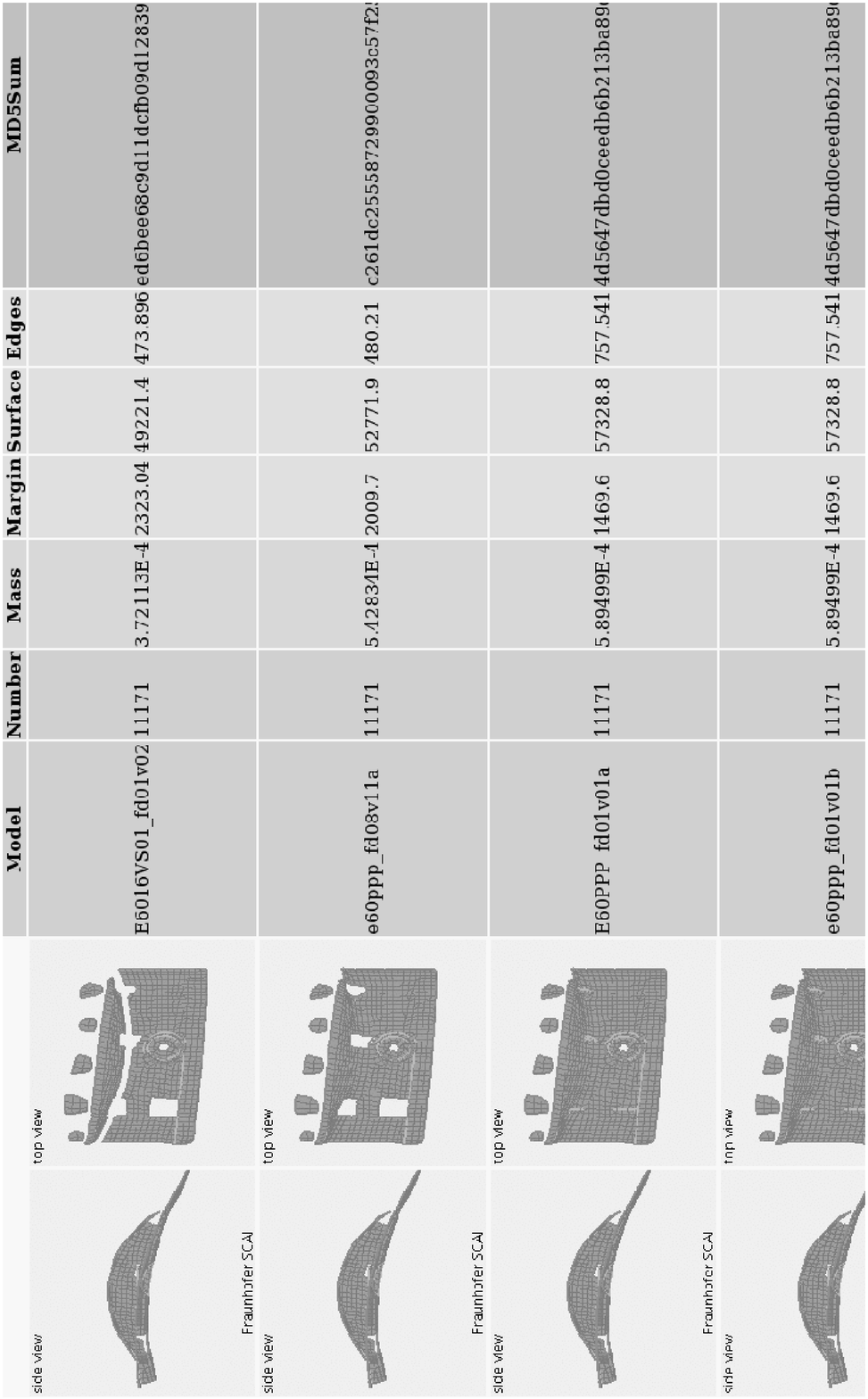}
\caption{Screenshot of SCAI--DM framework, here some variations of 
part no. 11171 with meta data and checksums. 
The database contains about 146.000 parts belonging to 134 different crash tests. 
Deleting all sub--mesh files with multiple MD5 checksums (right column) the 
database can be reduced by 93\% to 9900 different parts. }
\label{variations_of_part}
\end{figure}

\subsection*{c) Data cleaning and sorting --- clustering of parts}

The finite element models are subject to numerous kinds of modifications. 
During the engineering process in which a car model is improved with respect to its crash--worthiness 
a subset of parts is modified. In general the parts modified are the 
crash--relevant ones. Additional modifications follow the demands 
of other engineering disciplines, e.g. holes may be inserted into the parts in 
order to achieve a better drain of varnish during production. Such 
measures reduce crash-worthiness, which then again has to be improved 
by further modifications. 

However, not all parts are modified in all stages of car design. 
As unchanged parts cannot be responsible for deviations in the simulation 
results, such parts can be excluded from the analysis. In order to 
remove the parts that never have been modified MD5 checksums are created 
for all sub--mesh files, Fig.~\ref{variations_of_part} (right column). 
If the checksum of any part stays constant in any data set of 
interest the part was left unchanged and one single reference of the 
sub--mesh file is stored. Solely parts with more than one instance 
in the data base are included in the data mining queries.

\subsubsection*{SCAI--DM Data Base.}

After disassembling all parts are stored in the SCAI--DM framework along with their 
meta data, as shown in Fig.~\ref{variations_of_part}. Depending on the purpose parts and data can 
be displayed in any other combination.

\subsubsection{Avoiding Inconsistent Naming/Numbering.}

One bottle neck for data mining of the BMW data is the fact that text entries 
in the data management system are free text. Some agreements are complied with 
in the majority of cases. Repeatedly, however, re--naming and re--numbering of 
parts was encountered in the data, which showed that rules were not consequently 
followed. Therefore, to avoid irrelevant results from the analysis aimed at it is 
vital that all data entering the analysis stick to the same rules. The safest way to 
achieve correct data is to avoid the text entries in CAE--Bench altogether 
and use the FE descriptions as a basis. This again implies that an automatic 
method to identify parts has to be set up such that the use of 
part--numbers or --names coming from CAE--Bench is avoided.
The meta data calculated from the FE model can be the basis for part identification 
using cluster analysis. The clustering process divides a dataset into mutually exclusive 
groups such that the members of each group are as "close" as possible to one another, 
and different groups are as "far" as possible from one another, where distance is measured 
with respect to all available variables, see e.g. \cite{cluster_paper}. 

Each meta data property spans a 
new dimension in the similarity space. The meta data of a specific part are represented by a point in the multidimensional similarity 
space. Similar parts have similar meta data and form a cloud of adjacent points. 
Figure~\ref{hierarchical_clustering} (left) shows the idea of 
clustering of meta data in a schematic diagram: Two dimensions of the similarity space are shown. 
The meta data of the parts $L$ and $C$ form two clouds, in which a substructure indicates the presence 
of several modifications. Figure~\ref{hierarchical_clustering} (right) shows a clustering plot 
of the BMW data. The dots describe to two different parts with the 
same part number 11011 "Motorträger" (three clusters on the right) and 
"Schott\-blech Motorträger" (small cluster in the centre of the diagram). 
This is an example for a change in numbering of a part.

\subsection*{d) Similarity analysis and data reduction --- clustering of variants}

\begin{figure}[h]
\begin{minipage}[c]{0.38\textwidth}
\centering \includegraphics[width=1.0\textwidth,angle=0]{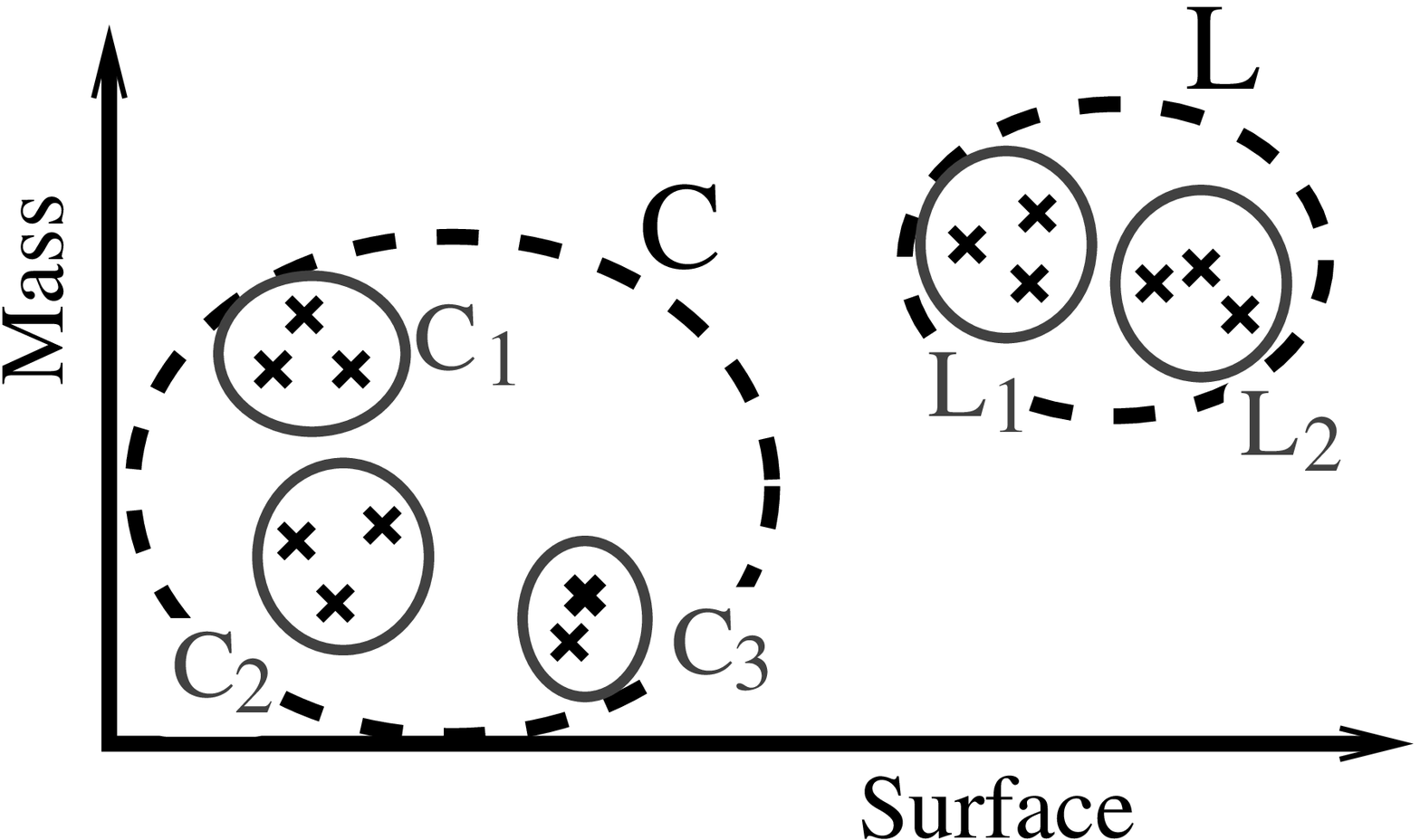}
\end{minipage}
\begin{minipage}[c]{0.62\textwidth}
\centering \includegraphics[width=3.7cm,angle=270]{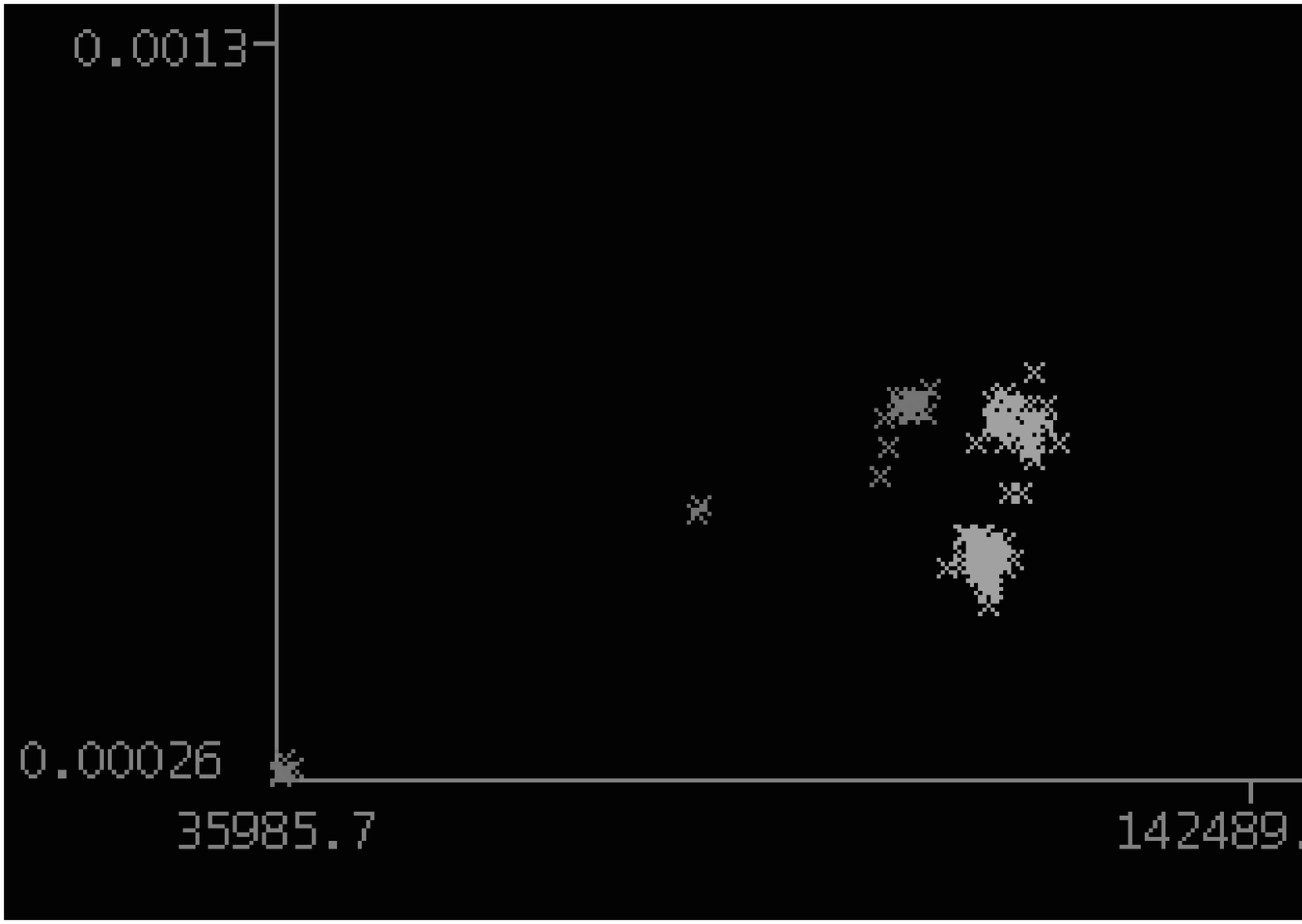}
\end{minipage}
\caption{Left: idea of hierarchical clustering in a schematic diagram. Two dimensions 
of the similarity space are shown. Right: meta data of real parts. The grey 
clusters on the right correspond to three different variants of the same part.}
\label{hierarchical_clustering}
\end{figure}

Differing checksums indicate that a sub--mesh was modified in some unknown way. 
Then negligible file modifications have to be distinguished from relevant changes 
such as modified shapes. In this framework for data mining the geometric meta data, 
as described above, serves as a similarity measure for the parts. Minor and major 
changes of the parts design will result in a hierarchical structure of clouds 
and sub--clouds, see Fig.~\ref{hierarchical_clustering} (left). 
Using hierarchical clustering a substructure can be found inside the clusters. 
Starting with $C_2$ as a reference, $C_1$ contains parts with a higher mass 
(caused by a higher thickness $d$ or density $\rho$ of the material) while the parts 
in $C_3$ result from geometrical modifications increasing the surface (e.g. caused 
by additional beadings for higher stiffness). In the clustering plot of BMW data, 
Fig.~\ref{hierarchical_clustering} (right), the light grey dots belong to three modifications 
of the same part, namely 11011 "Motorträger".
An example for typical modifications and their influences on the meta data can be 
seen in Fig.~\ref{variations_of_part}, where similar parts have been selected from the data base.

This clustering of parts in the meta data space in order to identify variants 
of designs is a time consuming task when all relevant parts and meta data are 
considered. An alternative method leading to similar results is to merge the 
meta data into a single similarity measure \cite{simsearch}. For the work 
presented in this paper a weighted sum of all the meta data has been employed. 
Then, if the weights are appropriately chosen, parts with the same similarity 
measure are similar in shape. This similarity measure serves as the main 
attribute for data mining, as described in section~\ref{part_ii}.

\subsection*{e) Evaluation and cleaning of crash result data}

For each crash simulation several values and curves, as well as images 
and movies are computed in order to evaluate the crash worthiness of 
this particular design. The bottle neck here is 
similar as before: the scripts that calculate the values stored in 
CAE--Bench can be altered at any time, such that the compatibility 
of the values has to be ensured before data mining can be attempted. 
No automatic approach could be developed to check this compatibility so far. 
In this work values whose scripts have been left unchanged for all simulations have been used for the 
DM analysis. This could, however, be a serious drawback of the method and other 
possibilities of ensuring reproducible values for the crash results 
have been discussed with BMW. In this paper the only result values 
analysed are intrusions. Intrusions measure the difference between the 
distances of two points inside the car (one FE node) before and after 
the crash test. 

In the last step of data preparation the data base is reordered. A table containing one line per crash test is formed: the name of the model,
the similarity values of the parts and the result values of interest.

\begin{figure}[h]
\centering \includegraphics[width=12.0cm,angle=0]{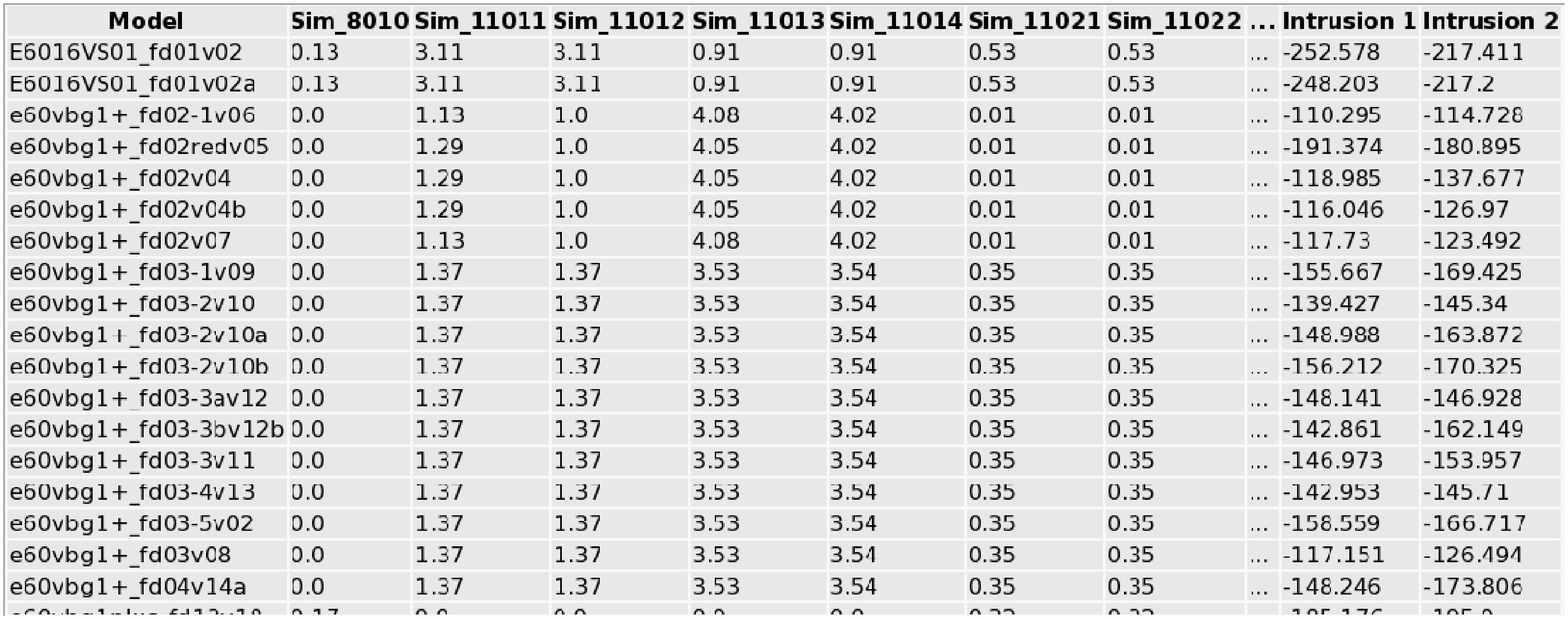}
\caption{Re--ordered table for data mining: each line contains one model 
with similarity values of relevant parts and serves as instance for analysis. 
Simulation results (intrusions) are attached and serve as classes.}
\label{simtable}
\end{figure}

\section{Datamining on similarity data}\label{part_ii}

The aim of this work is to evaluate the applicability of data mining methods 
for simulation data in engineering. As a result from a complex data preparation
procedure a table suitable for data mining can be achieved in which simulation 
data appears transformed into geometrical meta data.
This table (Fig.~\ref{simtable}) is written in Weka format, for which readily applicable 
data mining algorithms are available, see \cite{weka,weka2}.

\subsection{Attribute Selection}

An important step in data mining is the selection of those attributes that 
are relevant predictors before starting to build the model \cite{devaney}.
This is important because too many may be available when the full data set 
is encountered. Irrelevant information 
should be excluded from the data set \cite{blum}. Thus a feature selection algorithm can 
show which attributes have the strongest influence on the class. For the crash 
simulation data this information can be particularly valuable, as it reduces a 
vast amount of geometrical modifications to a small number of seemingly important ones.

\begin{figure}[h]
\centering \includegraphics[width=3.0cm,angle=270]{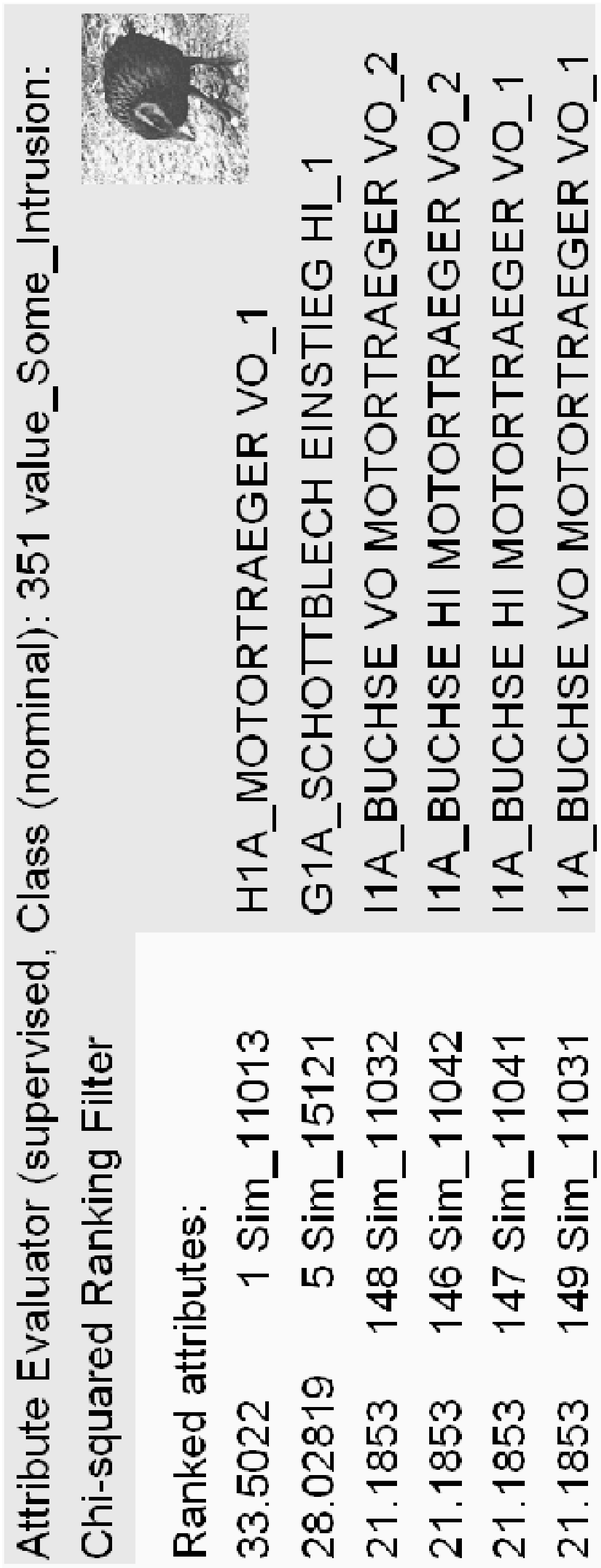}
\caption{Attribute selection with Weka: The six parts whose variations 
have most influence on the intrusion. BMW has confirmed the importance 
of these parts for the front crash simulated here.}
\label{attribute_selection}
\end{figure}

Employing an attribute selection algorithm on the crash data, e.g. ChiSquared 
in Weka, means that parts are ranked depending on the impact of the variation 
of their similarity measure on the intrusion of interest. In Figure~\ref{attribute_selection} one 
result of such a calculation is shown. The list of parts shows those 6 out of 1200 
whose variations have most influence on the intrusion. In this case the data basis is 30 
models---the portability therefore is likely to be rather limited.

\subsection{Decision Trees}

\begin{figure}[h]
\begin{minipage}[c]{0.5\textwidth}
\centering \includegraphics[width=4.4cm,angle=270]{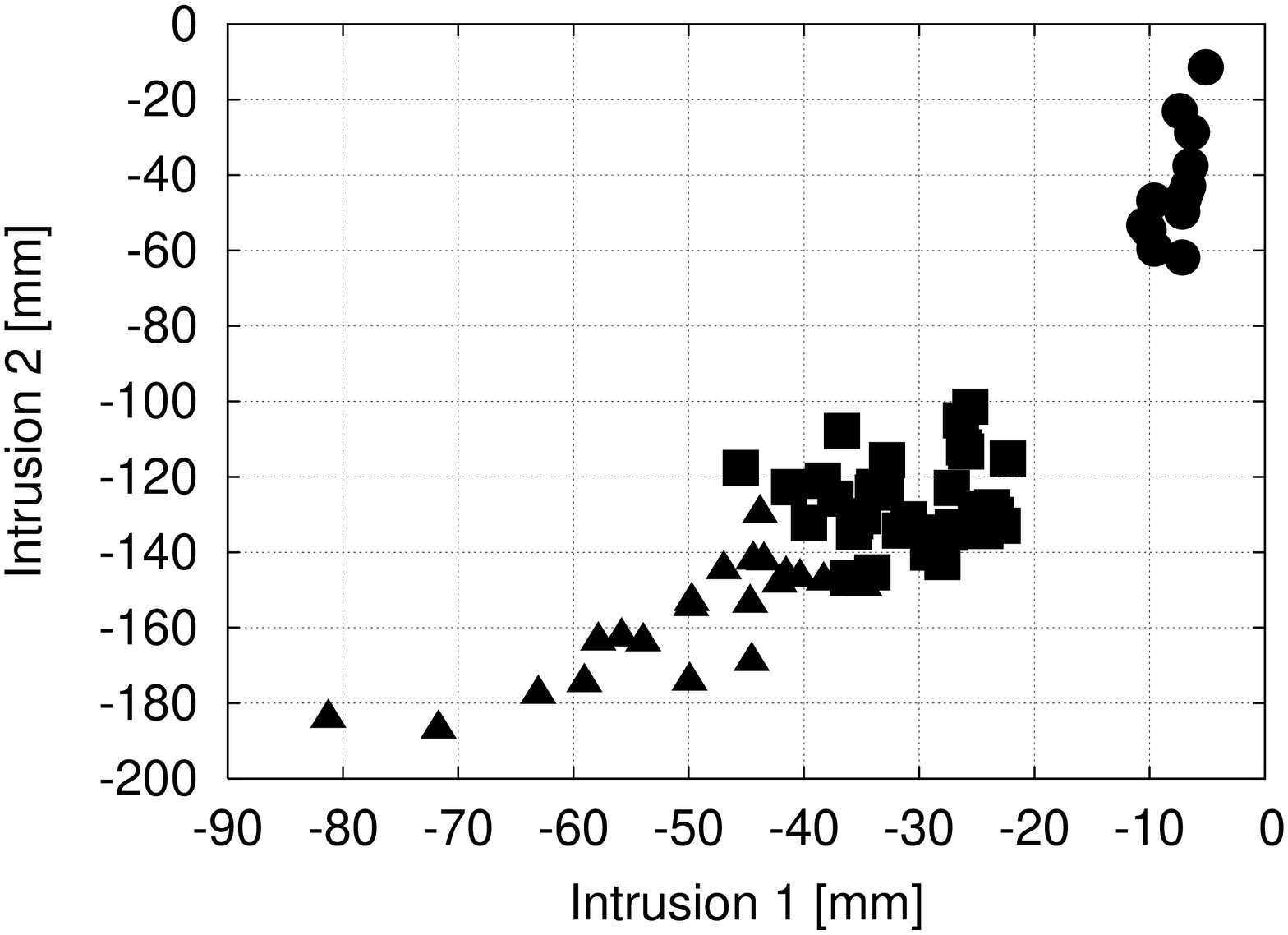}
\end{minipage}
\begin{minipage}[c]{0.5\textwidth}
\centering \includegraphics[width=4.4cm,angle=270]{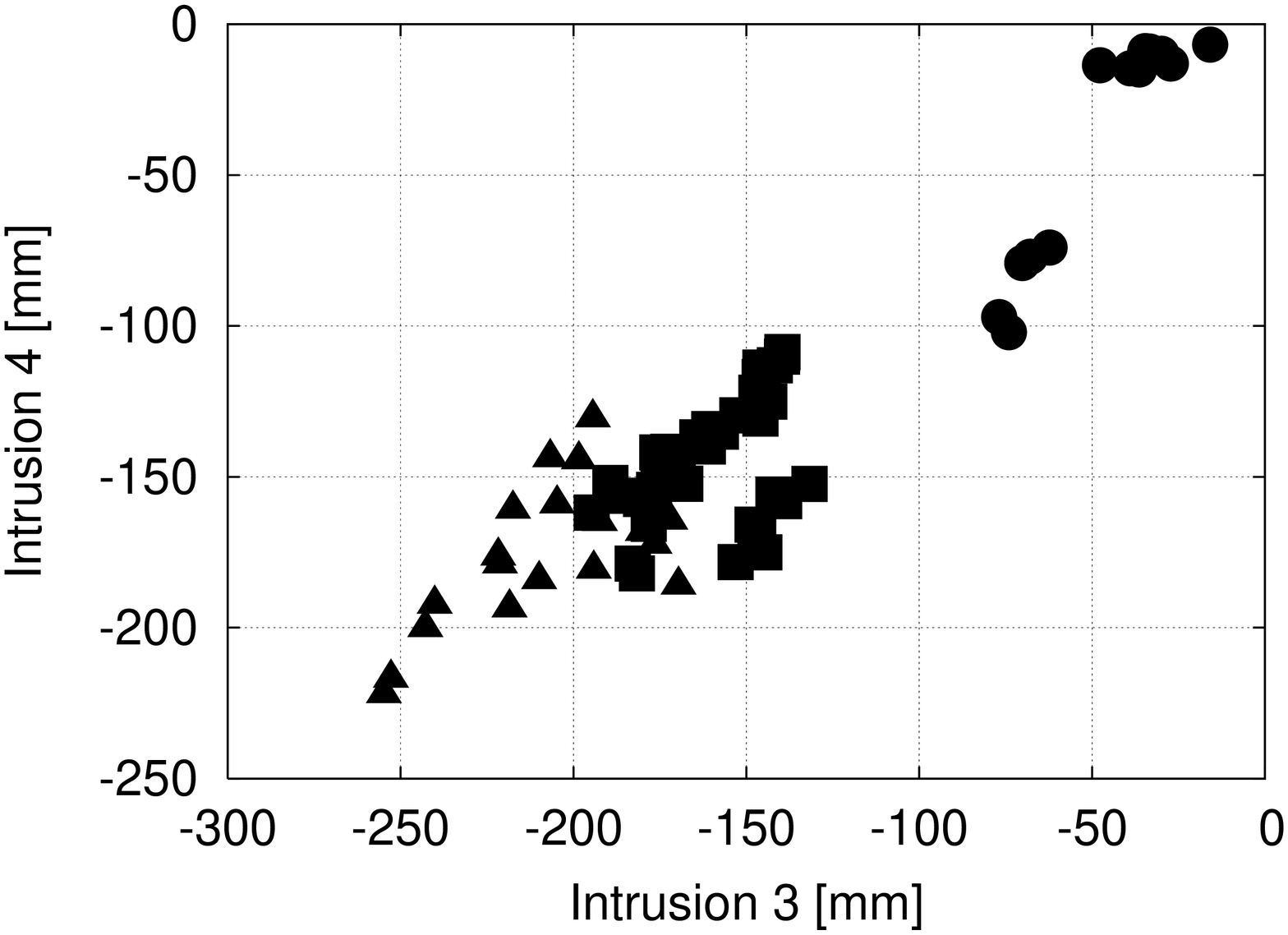}
\end{minipage}
\caption{Clustering of result values "Intrusion 1 \ldots 4" in order to be able to represent 
various aspects of crash behaviour with one single class value.}
\label{clustering_of_postprocess}
\end{figure}

Another method employed in order to demonstrate the possible outcome of 
data mining on simulation data is the decision tree method. Here a further 
step has been taken towards the achievement of results relevant to the application engineer.
In practice the engineer is very rarely interested in the behaviour of 
only one of his result values, instead he needs to get an understanding of the influences of his design modification on a range on values.
For this reason four result values were selected and clustered into 
three groups, one of which covers the most desired vehicle behaviour during crash.
The clustering of the instances into three groups is demonstrated in 
Figure~\ref{clustering_of_postprocess} for two of the result values. 
A clear grouping into "good" (circles), "medium" (squares) and "poor" (triangles) crash tests can be seen.

The membership of a model to these clusters is then used as "class" 
when the decision tree is built. As attributes the similarity measure---in this case 
again a weighted sum of all meta data---is employed.

\begin{figure}[h]
\centering \includegraphics[width=6.5cm,angle=270]{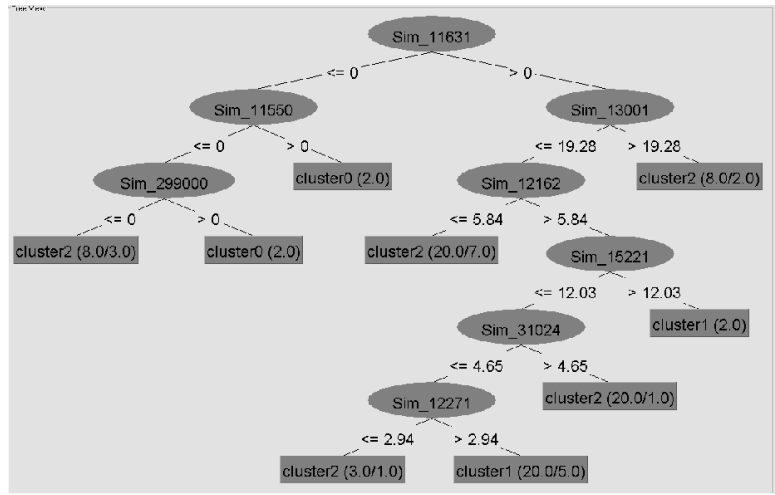}
\caption{Decision tree \texttt{weka.classifiers.trees.J48 -C 0.25 -M 2}. The 
existence of the part 11631 "Abstützung Lenksäule Unterteil" is the determining factor 
for the intrusions. If this part was integrated ($> 0$) further important parts in this data basis 
are two modifications of part 13001 "Bodenblech vorne" and part 12162 "Verbindung Längsträger".}
\label{dectree}
\end{figure}

An example for such a tree is shown in Figure~\ref{dectree}. 
The tree thus shows in which cluster a carmodel can be expected to lie depending on the geometrical version 
of the parts contained in the carmodel. These represent now the nodes of the tree.

For this example a data basis of 77 crashtests has been employed, which still 
is a rather small basis for rule building. However, these results are promising
because the similarity measure seems capable of adequately representing 
shape modifications and lead to meaningful results.

\subsection{DM Reports}

The results achieved within the SCAI-DM framework is imported 
into CAE--Bench in order to be accessible by other engineers at other times.
The reporting tool in CAE--Bench can include text and figures, such that a 
data mining report can be stored with the underlying input decks
in CAE--Bench. This closes the circle of the procedure shown in Fig.~\ref{procedure_for_data_mining}.

\section{Results}

The applicability of data mining on crash simulation data has been demonstrated in this work. 
A framework for data preparation has been developed. The computation and handling of 
meta data for similarity search has been studied in detail. The employed similarity measure 
has proved to be appropriate for detection of relevant changes in shape. 
The usability of the approach on data from an automotive application has been shown. 
Due to the limited amount of data available for this work conclusions are limited, but 
first significant results have been achieved on a test set of data. The next 
step aimed at is the integration of selected algorithms and data preparation tools into CAE--Bench. 
As soon as this has been accomplished the method needs to be validated on a more substantial  
data set, i.e. within the working environment of BMW. Then it will be feasible to judge whether the
original questions aimed at can be answered.

\section*{Acknowledgements}

We wish to thank T. Gholami from BMW AG for his support of this work.

\end{document}